\documentclass{aa}  

\newcommand{\mstar}{$M_{\star}\,$}
\defcitealias{Zheng20}{Z20}
\usepackage{graphicx}
\usepackage{txfonts}
\begin{document}

   \title{Shapes of galaxies hosting radio-loud AGNs with $z\leq1$}

   \author{X.~C. Zheng\inst{1}
   \and H.~J.~A. R\"{o}ttgering\inst{1}
   \and A.~van der Wel\inst{2,3}
   \and K.~Duncan\inst{4}
          }
   \institute{Leiden Observatory, Leiden University, PO Box 9513, NL-2300 RA Leiden, the Netherlands\\
              \email{zheng@strw.leidenuniv.nl}
              \and
              Sterrenkundig Observatorium, Department of Physics and Astronomy, Ghent University, Belgium
              \and
              Max-Planck Institut f\"{u}r Astronomie, K\"{o}nigstuhl 17, D-69117 Heidelberg, Germany
              \and
              Institute for Astronomy, Royal Observatory, Blackford Hill, Edinburgh, EH9 3HJ, UK
             }

   \date{}

 
  \abstract
   {{ Links between the properties of radio-loud active galactic nuclei (RLAGNs) and the morphology of their hosts may provide important clues for our understanding of how RLAGNs are triggered}.
   In this work, focusing on passive galaxies, we study the shape of the hosts of RLAGNs selected from the {\it Karl G. Jansky} Very Large Array Cosmic Evolution Survey (VLA-COSMOS) 3GHz Large Project, and compare them with previous results based on the first data release (DR1) of the LOFAR Two-Metre Sky Survey (LoTSS).
   We find that, at redshifts of between 0.6 and 1, high-luminosity ($L_{1.4\rm\,GHz}\gtrsim10^{24}\rm\,W\,Hz^{-1}$) RLAGNs have a wider range of optical projected axis ratios than their low-redshift counterparts, which are essentially all found in round galaxies with axis ratios of higher than 0.7.
   We construct control samples and show that although the hosts of high-redshift RLAGNs with the highest luminosities still have a rounder shape compared with the non-RLAGNs, they on average have a smaller axis ratio (more elongated) than the local RLAGNs with similar stellar masses and radio luminosities.  
   This evolution can be interpreted as a byproduct of radio luminosity evolution, namely that galaxies at fixed stellar mass are more radio luminous at high redshifts: artificially increasing the radio luminosities of local galaxies ($z\leq$0.3) by a factor of 2 to 4 can remove the observed evolution of the axis ratio distribution. 
   If this interpretation is correct then the implication is that the link between AGN radio luminosity and host galaxy shape is similar at $z\simeq1$ to in the present-day Universe.
   }


   \maketitle
%

\section{Introduction}

The coevolution of galaxies and supermassive black holes (SMBHs) is an important topic in astrophysics.
Evidence suggests that the feedback from active SMBHs in the centres of galaxies, namely active galactic nuclei (AGNs), plays an important role in modulating and quenching star-forming activity \citep[e.g.][]{Gebhardt00,Greene06,Heckman14}.
In this picture, the energy ejected by the radio jet of radio-loud active galactic nuclei (RLAGNs) is crucial for heating the intergalactic medium and preventing further star formation from the cold gas in galaxies \citep{Best06,McNamara07,Fabian12}. 
However, the details of the triggering of radio jets are not yet clear.

One of the important factors affecting the jet-launching process of SMBHs is the black hole spin, which is thought to be responsible for the wide range of radio loudness in AGNs \citep[see e.g.][]{Wilson95,Fanidakis11}.
As described by the Blandford-Znajek mechanism \citep{Blandford77}, which is the most popular analytical jet-launching model, a rotating SMBH can produce a highly collimated relativistic jet via a strong magnetic field.
Therefore, the spinning up of SMBHs is a key issue, and this spin can be increased by either a major merger or a series of accretion events, possibly induced by minor mergers \citep{Sikora07,Fanidakis11}.
The two spin-up paths also lead to different shapes in the resulting galaxies, which connects RLAGNs with the morphology of their hosts.

It has long been known that powerful RLAGNs are usually observed in massive elliptical galaxies \citep[e.g.][]{Condon78,Balick82,Best05b,Brown11,Vaddi16}.
Recent works \citep[hereafter \citetalias{Zheng20}]{Barisic19,Zheng20} also show that galaxies with a larger axis ratio have a higher chance of hosting a high-luminosity RLAGN.
This trend can be explained by major mergers, because they can produce massive elliptical galaxies with both a large optical projected axis ratio (the ratio of the minor axis to the major axis, hereafter $q$) and a fast-spinning SMBH.
On the other hand, recent work \citep{Sadler14,Tadhunter16,Pierce19,Wang16,Wang19} on local low-luminosity RLAGNs showed that secular processes (e.g. disc instabilities or bar-related processes) may also lead to RLAGNs in disc-like galaxies with an elongated shape.
This difference in host-galaxy shape between high- and low-luminosity RLAGNs below redshift 0.3 was clearly seen by \citetalias{Zheng20} from high-quality data from the LOFAR Two-metre Sky Survey \citep[LoTSS;][]{Shimwell19} and the Sloan Digital Sky Survey \citep[SDSS;][]{Strauss02}. 
However, we have little knowledge about RLAGNs at higher redshift because of the difficulty in obtaining a large amount of reliable morphological and radio data.

\begin{figure*}
    \centering
    \includegraphics[width=\linewidth]{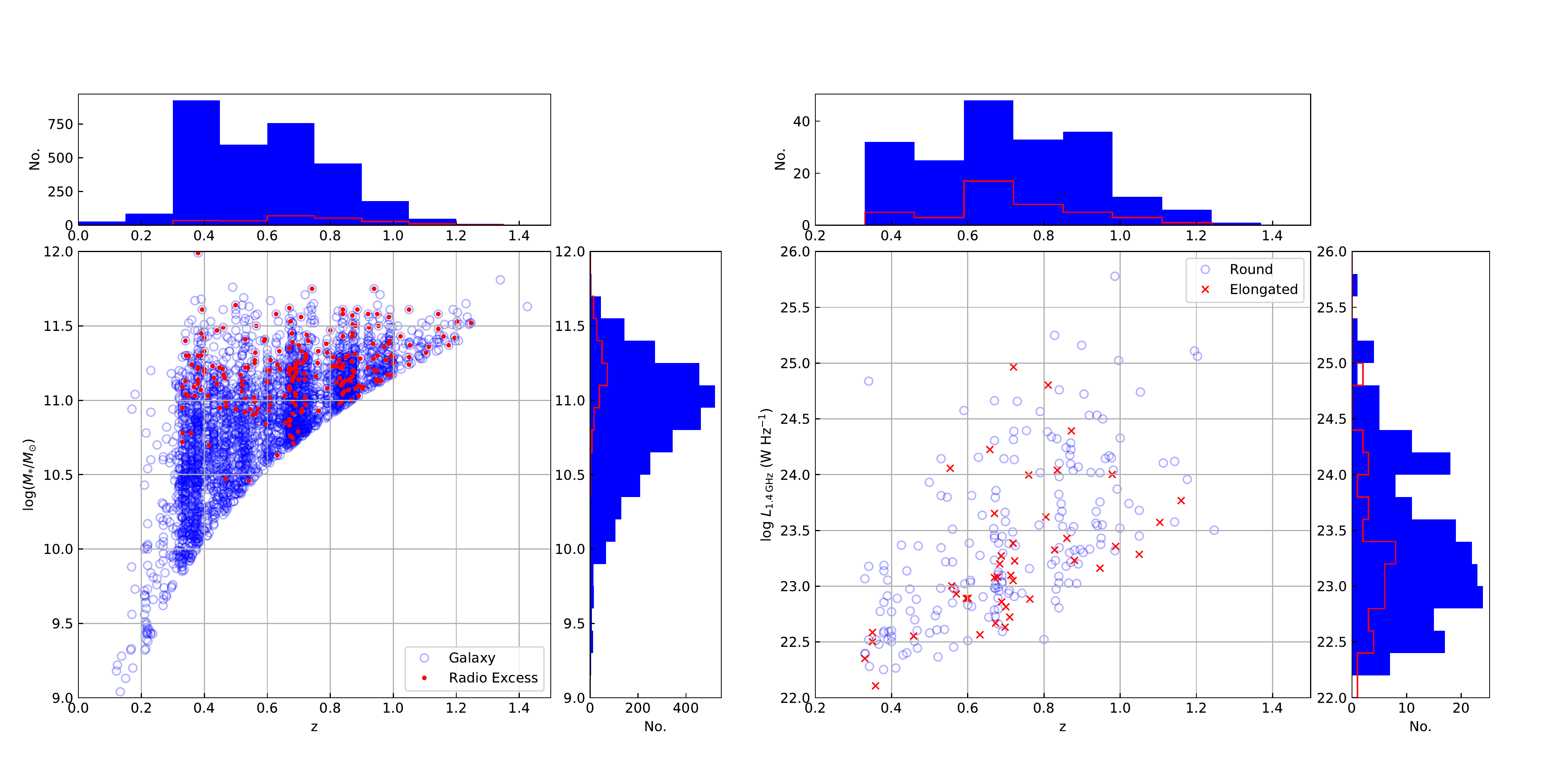}
    \caption{(Left) Distribution of \mstar and redshift of the galaxies in this work.
    Blue circles and histograms represent passive galaxies and their \mstar or redshift distributions.
    Red dots and the red histograms are for the RLAGNs.
    (Right) Distribution of rest-frame 1.4 GHz AGN radio luminosity and redshift of the RLAGNs used in this work.
    Blue circles and histograms represent round ($q>0.6$) galaxies and their radio luminosity or redshift distributions, while red crosses and histograms are for the elongated ($q\leq0.6$) galaxies. 
    }
    \label{fig:mldist}
\end{figure*}
In this work, we study RLAGNs with $0.3<z\leq1$ using morphological data based on high-resolution images from the Hubble Space Telescope (HST) and radio data from the Very Large Array (VLA).
The article is structured as follows.
Section \ref{sec:data} describes the data and sample selection in this work.
The analyses of the data are shown in Sect. \ref{sec:analysis}.
We then summarise our results and discuss their physical implications in Sect.\ref{sec:conclusion}.
We adopt a standard cosmology with $H_0=70\rm\,km\,s^{-1}\,Mpc^{-1}$, $\Omega_{\rm M}=0.3$, $\Omega_{\Gamma}=0.7,$ and an initial mass function (IMF) from \citet{Chabrier03} throughout this paper.
%

\section{Data and sample selection}\label{sec:data}

Our radio data are based on the {\it Karl G. Jansky} Very Large Array Cosmic Evolution Survey (VLA-COSMOS) 3GHz Large project \citep{Smolcic17a}.
The 3GHz Large project covers an area of $\sim$2.6 square degrees, enclosing the 2 square degree COSMOS field with 384 hours of observations, and reaches a mean rms of $\sim$2.3$\mu$Jy/beam and a resolution of 0.75\arcsec.
This project resulted in a source catalogue \citep{Smolcic17a} containing 10830 radio sources above 5$\sigma$ with an astrometric accuracy of about 0.01\arcsec\,at the bright end.
Most ($\sim93\%$) of the radio sources in this catalogue also have multiwavelength information from infrared to X-ray, which helps in determination of the sources properties, and in separating star-forming galaxies from RLAGNs.

The COSMOS2015 catalogue \citep{Laigle16} provides optical to near-infrared photometry for 1182108 sources in 30 bands over an area of 2 square degrees. 
The catalogue also lists photometric redshifts with a high precision of $\sigma_{\Delta z/(1+z)}$ of better than 0.01 at $z\leq 1$\citep{Laigle16} based on the extensive photometry, and provides stellar mass estimates derived using the \citet{BC03} model with a \citet{Chabrier03} IMF.
A total of 7729 radio sources have a counterpart in the COSMOS2015 catalogue with a false positive rate of $\sim2\%$ \citep{Smolcic17b}.
Hereafter, we use the redshifts and stellar masses listed in the COSMOS2015 catalogue for the sources in this work.

We adopt the morphological measurements from the Advanced Camera for Surveys General Catalog \citep[ACS-GC; ][]{Griffith12}, a photometric and morphological database of data from the Advanced Camera for Surveys (ACS) instrument \citep{Clampin02} on the HST.
The photometric and morphological information for sources in the COSMOS field in the ACS-GC are computed with {\tt SEXTRACTER} and {\tt GALFIT} using the 2028s F814W images described in \citet{Koekemoer07}, which have a pixel scale of $\sim$0.05\arcsec/ pixel.
This catalogue contains 304688 sources within a $\sim$1.8 square degree area in the COSMOS field.

Because there is no existing direct cross-matching information between ACS-GC and the other two catalogues, we use the $i-$band-selected catalogue described in \citet{Capak07} to guide cross-matching.
This is because both the ACS-GC and the COSMOS2015 list the counterpart ID in the $i-$band-selected catalogue for their sources when available.
We find that 232682 sources in the COSMOS2015 catalogue and ACS-GC can be matched through the same IDs in the $i-$band selected catalogue.
Of these matched sources, 99.9\% have a positional difference of less than 0.8\arcsec\ .
We then apply this positional difference as a searching radius to cross-match the ACS-GC against the COSMOS2015 catalogue directly to find more sources without counterparts in the $i-$band-selected catalogue.
Finally, we obtain a sample containing 281940 sources, of which 5064 sources also have a radio counterpart in the 3GHz catalogue.
\begin{figure*}
    \centering
    \includegraphics[width=\linewidth]{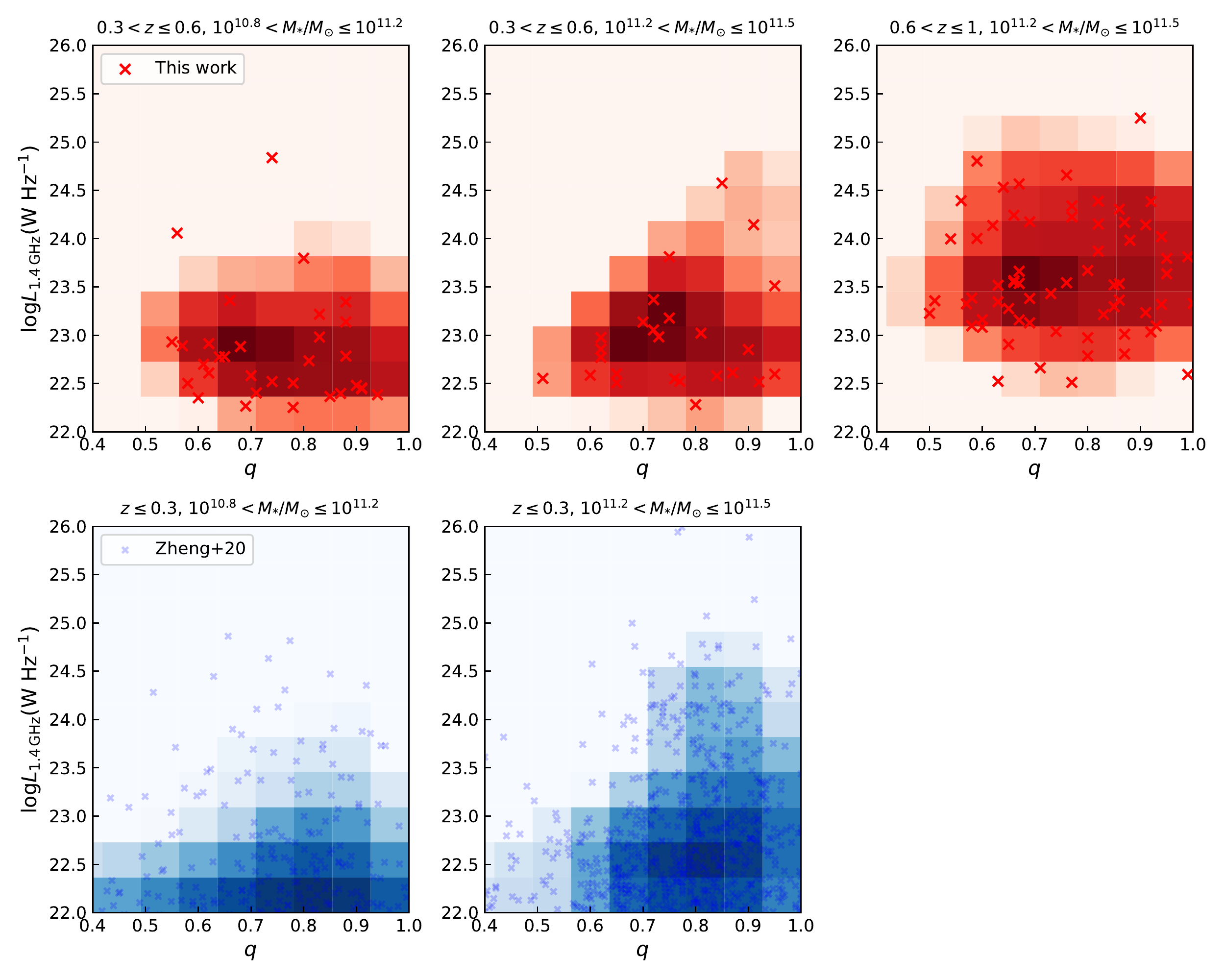}
    \caption{Distribution of RLAGNs with different redshifts at fixed \mstar in the $L_{1.4\,\rm GHz}$--$q$ plane.
    The top three panels show the distributions of RLAGNs selected in this work, while the bottom panels show the distributions of RLAGNs presented by \citet{Zheng20} and based on LoTSS DR1. 
    The redshift and \mstar ranges are noted at the top of each panel.
    Smoothed results based on KDE are shown in the background as 2D histograms.}
    \label{fig:Lvsq}
\end{figure*} 

To construct a useful sample for the following analysis, it is important to (1) select the passive galaxies, (2) find out the mass completeness limit, and (3) distinguish the real RLAGNs from the star-forming galaxies.
For simplicity, we limit our analysis to the passive galaxies; active star formation would contribute to the radio emission, but star-forming galaxies tend to have very different properties compared with passive galaxies and are less likely to host a RLAGN \citep{Janssen12}. 
In the COSMOS2015 catalogue, the star and galaxy classifications (stored in flag `OType') were obtained by comparing the spectral energy distribution (SED) fitting results using galaxies and stellar templates \citep{Laigle16}.
The passive galaxies were then identified using the ${\rm NUV}-r/r-J$ colour--colour criterion \citep[][flagged as `Cl=0' in the COSMOS2015 catalogue]{Williams09}.
Benefiting from the comprehensive results in the COSMOS2015 catalogue, we are able to select passive galaxies with the flags `OType'=0 and `Cl'=0.

Stellar mass (\mstar) is one of the most important factors determining both the fraction of RLAGNs and the morphology of galaxies \citep[see e.g.][]{Sabater19,Cappellari16}.
Therefore, the next step is to derive the mass completeness limit as a function of redshift in our sample; otherwise, the analyses using low-mass galaxies would be limited to the brightest ones, which might cause bias in the final results.
The completeness limit in our work is mainly constrained by ACS-GC.
As suggested in \citet{Griffith12}, ACS-GC achieves a completeness of 90\% at F814W $\approx23$ and 85\% at F814W $\approx23.5$.
Following a similar method to that described in \citet{Chang15}, we select sources with F814W$\leq$23.5 in ACS-GC, and then find the most massive galaxies with $23<{\rm F814W}\leq23.5$ in different redshift bins to determine the \mstar\, limit as a function of redshift, $z$.
As such, we find that the \mstar\, limit, $M_{\star,\rm lim}(z)$, can be represented by $M_{\star,\rm lim}(z)=2.79 {\rm log}(z/0.1)+8.38$ at $z<1.5$.
We therefore do not take objects below the completeness
limit into account  in order to avoid the bias from the low-mass galaxies.

The final step is to select the RLAGNs in the 3GHz catalogues, as a large fraction of the radio sources arise from star-forming activity in the galaxies instead of radio jets from central SMBHs.
Here we choose galaxies flagged as `radio excess' sources in the 3GHz catalogue to be the RLAGN sample.
\citet{Smolcic17b} obtained the star-formation rates (SFRs) based on the total infrared luminosity derived from SED fitting \citep{Delvecchio17}.
The SFRs are then converted to radio luminosities \citep{Condon92} contributed by star-forming activity in the galaxies.
The radio excess sources are those with  significantly
greater radio  luminosities than expected from the SFRs.

It should be noted that the concern that removing star-forming galaxies might also preferentially remove radiative-mode RLAGNs because they tend to be young galaxies \citep{Best12} is not necessary in this work.
Without removing the star-forming galaxies, only two RLAGNs with significant radiative emission in other wavelengths will be added to the final sample, which is consistent with the fact that the radiative-mode RLAGNs only make up a small fraction of the sample below $L_{1.4\rm GHz}\simeq10^{25}\,\rm W\,Hz^{-1}$.
However, including star-forming galaxies will enlarge the final non-RLAGN galaxy sample by over a factor of two.
These star-forming galaxies have significantly different properties, such as gas abundance, mass-to-light ratio, and metallicity, compared with the passive galaxies and RLAGN hosts.
We therefore exclude them to avoid potentially introducing biases and increasing the complexity of the following analyses.

Applying these steps means that we obtain a sample containing 3093 passive galaxies, of which 234 sources host a RLAGN.
The distribution of \mstar, radio luminosity, and redshift of the sample is presented in Fig.\ref{fig:mldist}.
It shows that the redshift of galaxies in our work ranges from 0.1 to 1.5, and is mostly between 0.3 and 1.
The majority of the galaxies have \mstar in the range $10^{10}\,M_{\odot}$ to $10^{11.5}\,M_{\odot}$, while galaxies with a RLAGN are nearly all more massive than $10^{10.5}\,M_{\odot}$.
The rest-frame 1.4 GHz radio luminosity of the RLAGNs in this work is taken from the rest-frame 1.4 GHz luminosity listed in the 3GHz catalogue; most have a $L_{1.4\,\rm GHz}$ ranging from $10^{22}$ to $10^{25}\,\rm W\,Hz^{-1}$.
\begin{figure*}
    \centering
    \includegraphics[width=\linewidth]{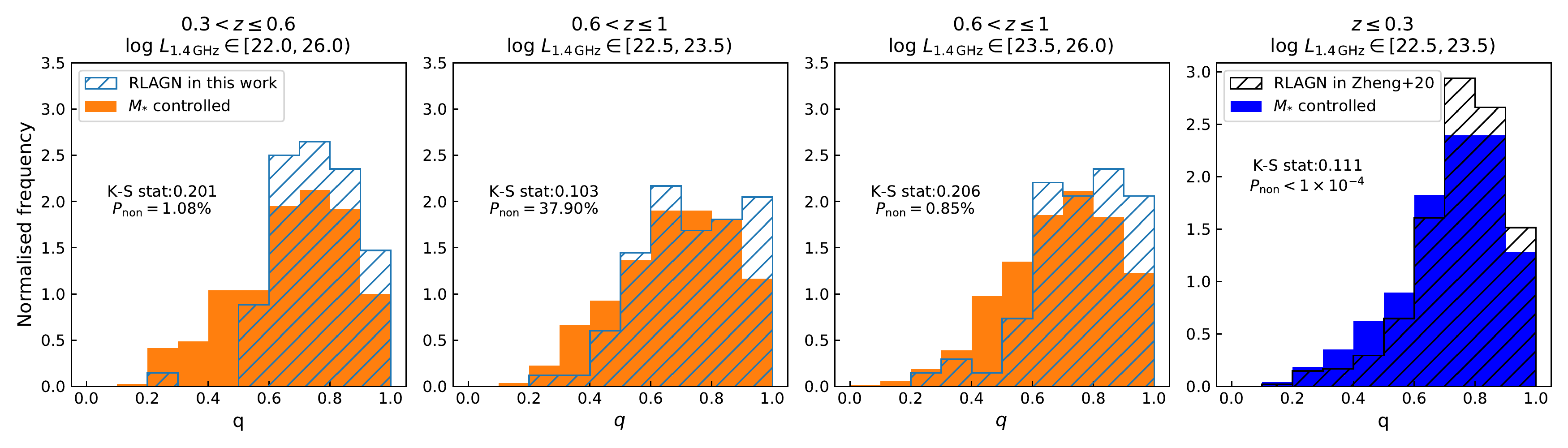}
    \caption{Projected axis ratio distributions of RLAGNs and the non-RLAGN control sample.
    The first three panels show the distribution for sources within redshift 0.3 to 1 selected in this work.
    The last panel, for comparison, shows sources with redshift of less than 0.3 based on \citetalias{Zheng20}.
    The filled histograms represent the normalised frequency distributions for non-RLAGN galaxies, while the over-plotted blue hatched histograms represent the hosts of RLAGNs.
    The radio luminosity and redshift range of the RLAGN samples are noted at the top of the panels.
    The $p$-values derived from the median of 100 K-S tests are listed in the panels.
    }
    \label{fig:qdist}
\end{figure*}

For comparison, we use the sample described in \citetalias{Zheng20} as a local sample. 
This is based on data from LoTSS DR1 \citep{Shimwell19,Williams19} and SDSS DR7 \citep{Strauss02} and uses the same sample selection as that described above: passive galaxies only, and the mass limit as calculated in \citet{Chang15}.
The RLAGNs in \citetalias{Zheng20} are classified based on the comprehensive criteria in \citet{Sabater19}, which used the combined information from radio luminosity, infrared colours, emission lines, and $D_{4000}$.
Although this classification is not the same as in this work, the difference should mainly influence the selection of low-power RLAGNs, and therefore would not impact the following analyses, as they focus on RLAGNs with $L_{150\rm\,MHz}>10^{23}\rm W\,Hz^{-1}$ (equivalent to $L_{1.4\rm\,GHz}>10^{22.3}\rm W\,Hz^{-1}$ assuming a canonical spectral index of 0.7). 
As a result, the low-redshift sample contains 15934 passive galaxies with redshift $z\leq0.3$, of which 1912 sources host a RLAGN with $L_{150\rm\,MHz}>10^{21}\rm W\,Hz^{-1}$.
The axis ratios of these low-redshift sources are taken from the SDSS pipeline and the reliability of these was discussed by \citetalias{Zheng20}.

A potential caveat is that the 3GHz catalogue is expected to contain more flat-spectrum radio sources. 
However, based on the estimation in \citet{Smolcic17a}, sources in the 3GHz catalogue have an expected median radio spectral index of about -0.7, which is close to the canonical value \citep{Condon02}. 
Similarly, \citet{Sabater19} also found a median spectral index (-0.63) close to -0.7 for RLAGNs in LoTSS DR1.
Because of the similarity in spectral indices in these two datasets and the rarity of flat-spectrum sources, we conclude that the possible flat-spectrum bias does not influence our following analyses statistically.

\section{Data analysis}\label{sec:analysis}
\subsection{Radio luminosity versus axis ratio}\label{sec:Lvsq}
At low redshift, the hosts of RLAGNs with different radio power have different distributions of morphological features \citepalias{Zheng20}. 
High-luminosity RLAGNs always have a large axis ratio \citepalias[see Fig. 3 in][]{Zheng20} at fixed \mstar.
This seems to change at high redshift.
Figure \ref{fig:mldist} shows the difference between the distributions of `round' ($q>0.6$) and `elongated' ($q\leq 0.6$) RLAGNs.
The fraction of elongated RLAGNs with $L_{1.4\,\rm GHz}\gtrsim10^{23}\,\rm W\,Hz^{-1}$ seems to be larger at $z>0.6$.

To see the change in the relation between radio power and galaxy shape more clearly, we can compare the distribution of RLAGNs with different redshifts in the $L_{1.4\rm\,GHz}-q$ plane.
We first constrain the analysis to RLAGNs within two small ranges of \mstar ($10^{10.8}-10^{11.2}\rm\,M_{\odot}$ and $10^{11.2}-10^{11.5}\rm\,M_{\odot}$), { so that the bias due to the difference in the \mstar\,distribution can be ignored for RLAGNs with redshift of less than 1} (see Fig.\ref{fig:mldist}).
We then split the sources into three subsamples according to their redshifts.
The distributions of the three subsamples after smoothing using kernel-density estimation (KDE) are shown in the upper panels of Fig.\ref{fig:Lvsq}.
For comparison, in the bottom panels, we also show the distributions of local RLAGNs from \citetalias{Zheng20} with similar stellar mass ranges.

The distributions of the three subsamples are quite different.
Although the source numbers are limited for the two lower redshift subsamples ($0.3<z\leq0.6$), their distributions seem similar to those of their local counterparts:
(1) low-mass RLAGNs show no correlation between radio power and $q$;
(2) high-mass RLAGNs seem to be rounder (larger axis ratio) at higher radio luminosity.
However, at higher redshift ($0.6<z\leq1$), with a similar mass range to the low-redshift and local high-mass subsamples, a significant fraction of high-power RLAGNs have a relatively small ($q<0.7$) axis ratio.
The link between $L_{1.4\rm\,GHz}$ and $q$ seems to disappear at high redshift.

We argue that this change in the shape distribution of RLAGN at high redshift is not caused by the difference in spatial resolution of the two surveys.
On the one hand, the excellent resolution of ACS-HST ensures that the morphological measurements are reliable for such galaxies at $z\leq1$, as the errors in $q$ are mostly smaller than 0.02 and do not vary with redshift in this work.
On the other hand, while ACS-HST has a better spatial resolution (in kpc) than SDSS in this work \footnote{The typical angular resolution of HST is about 0.1$\arcsec$, which corresponds to a spatial resolution of about 0.7 kpc at redshift 0.7. Taking a typical angular resolution of 1.2$\arcsec$ in SDSS, we have a spatial resolution of about 3.6 kpc at redshift 0.15.}, we find that the axis ratio measurement at fixed \mstar in the SDSS does not vary significantly with redshift (see Appendix in \citetalias{Zheng20} and Fig. \ref{fig:sdssab}), which means that the difference in the spatial resolution does not significantly bias our results.
This can also be inferred from the results in Fig. \ref{fig:Lvsq}, where the subsamples with redshift 0.3 to 0.6, which should have the best spatial resolution in this work, have $q$ distributions similar to the local SDSS-based subsamples rather than the higher redshift subsample.
Moreover, the integrated magnitudes of the 2D Sersic models from \citet{Griffith12} are approximately equal to ground-based magnitudes at a similar wavelength, indicating little or no low-surface brightness emission is missed by HST for these relatively bright galaxies \citep[e.g.][]{vdWel21}.

The difference in the rest-frame wavelengths in which $q$s are measured is also not important in the analysis.
Considering that the $g$ band filter of SDSS works at wavelengths close to the rest-frame wavelengths at which the high-redshift galaxies are observed, we compare the $q$s measured in $g$ band and those measured in $r$ band (used in \citetalias{Zheng20}) from the SDSS pipeline for galaxies used in \citetalias{Zheng20}.
The average ratio between the two $q$ measurements is $q_{g}/q_r=0.99$, which indicates that the $q$s of galaxies measured in these wavelengths are broadly consistent.

\subsection{RLAGN and non-RLAGN galaxies}\label{sec:control}

To study the triggering  of RLAGNs, it is also important to investigate the difference between RLAGN hosts and non-RLAGN galaxies.
The different \mstar distributions of these two populations could be the most important source of systematic error, because more massive galaxies have a higher probability of hosting a RLAGN and are also more likely to have a rounder shape \citep{Chang13,Sabater19}.
Therefore, we used a control sample analysis similar to \citetalias{Zheng20} to compare the shapes of RLAGN hosts and non-RLAGN galaxies directly.
\begin{figure*}
    \centering
    \includegraphics[width=\linewidth]{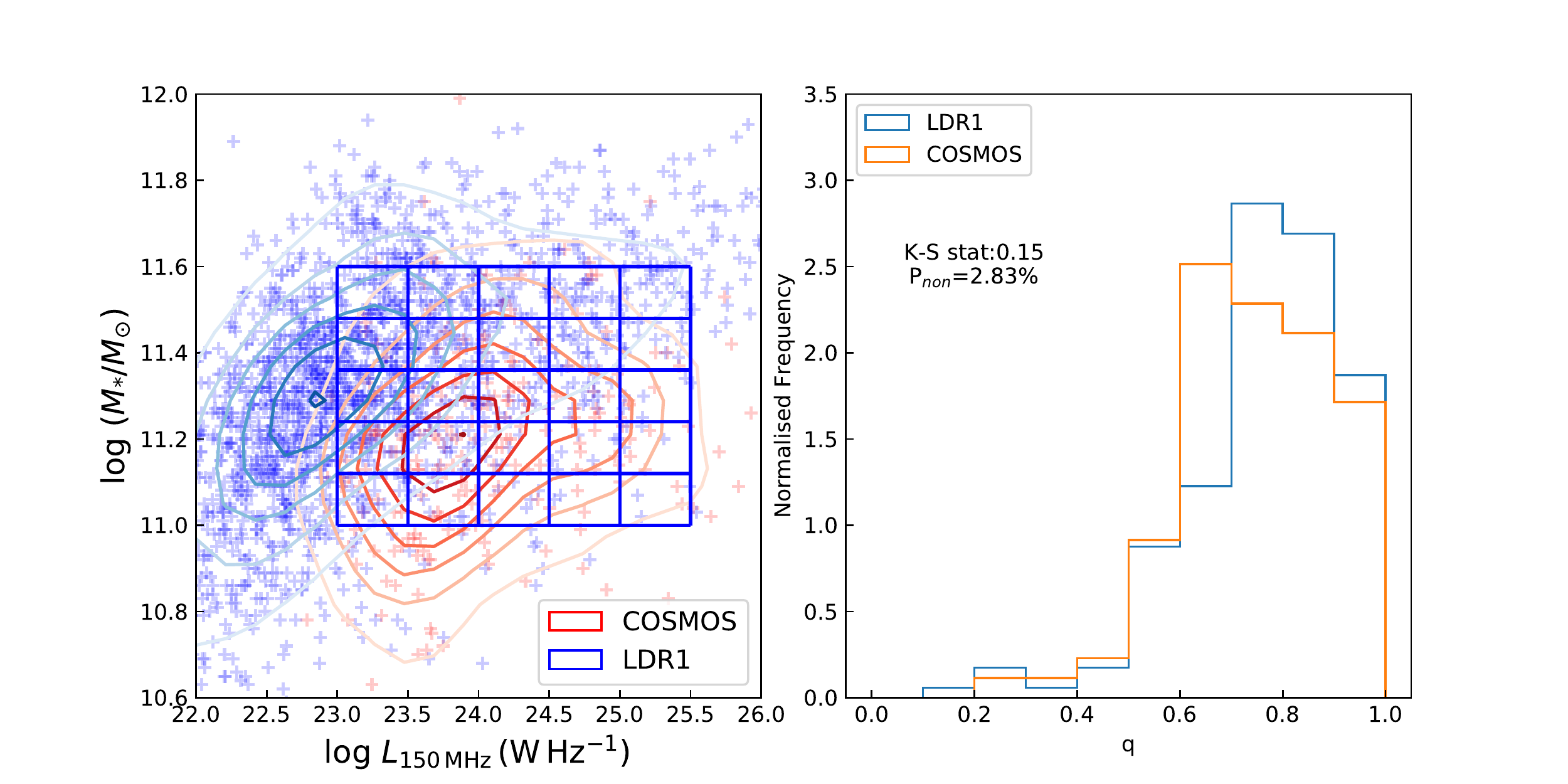}
    \caption{(Left) Stellar mass and 150 MHz radio luminosity of RLAGNs in this work and in \citetalias{Zheng20}.
    The red crosses and contours represent RLAGNs in the COSMOS field used in this work, while the blue crosses and contours represent RLAGNs from the LoTSS DR1 used in \citetalias{Zheng20}.
    The blue rectangle marks the sampling region and the bins used to construct the control sample.
    (Right) Projected axis ratio distributions of RLAGNs in this work and in the low-redshift control sample based on RLAGNs in the LoTSS DR1.
    The orange histogram denotes the normalised frequency distribution of RLAGNs in this work, while the blue histogram denotes the distribution of RLAGNs in the low-redshift control sample.
    The K-S result listed here is based on the median of 1000 tests.
    }
    \label{fig:compare}
\end{figure*}

Firstly, we separated the RLAGNs into three subsamples based on their redshifts and radio luminosities, as noted in Fig.\ref{fig:qdist}.
The range of radio luminosities here is chosen for the purpose of ensuring sufficient sample statistics (about 70) in each subsample.
We then obtained the \mstar\,distribution of each subsample; this is given by the source numbers within six logarithmic equidistant bins between $10^{10.6}M_{\odot}$ and $10^{11.6}M_{\odot}$.
Next, we randomly took 800\footnote{About ten times the source number of the largest RLAGN subsample.} non-RLAGN galaxies in the same redshift range, with a \mstar\, distribution similar to that of RLAGNs, as a representative \mstar-controlled sample.
In this way, any influence from the difference in \mstar distributions should be eliminated, and the $q$ distribution differences of RLAGN and non-RLAGN sources should only be driven by their different radio properties. 
We then performed K-S tests to compare each RLAGN subsample with the corresponding control sample.
This process was repeated 100 times to reduce the stochastic error.
We show the $q$ distribution with the median K-S statistic from 100 tests for all subsamples and their control samples in the first three panels of Fig.\ref{fig:qdist}. 
For comparison, in the last panel we also show the $q$ distribution from a similar analysis for local sources by \citetalias{Zheng20}.

Inspection of Fig.\ref{fig:qdist} shows that the low-redshift RLAGN subsample has a larger fraction of sources with large $q$ than the non-RLAGN control sample, and the K-S test gives a $p$-value smaller than 5\%.
This means that the RLAGNs are significantly rounder than non-RLAGNs within redshifts 0.3 to 0.6, which is similar to the results for local RLAGNs \citepalias{Zheng20}.

This trend changes at higher redshift ($0.6<z\leq1$).
The very high luminosity ($L_{1.4\rm\,GHz}\geq10^{23.5}\rm\,W\,Hz^{-1}$) RLAGNs still have a $q$ distribution that is significantly different from that of the control sample, and is similar to the results for low-redshift RLAGNs.
However, the K-S test for RLAGNs with $L_{1.4\rm\,GHz}$ between $10^{22.5}$ and $10^{23.5}\rm\,W\,Hz^{-1}$ gives a $p$-value much larger than 5\%, and therefore we cannot reject the null hypothesis.
This result means that the RLAGNs in this subsample and in the non-RLAGN control sample do not show significant differences.
Within the same luminosity range, as shown in the last panel of Fig. \ref{fig:qdist}, local RLAGNs are typically rounder than non-RLAGNs.
Therefore, the $q$ distribution of RLAGNs with $L_{1.4\rm\,GHz}$ between $10^{22.5}$ and $10^{23.5}\rm\,W\,Hz^{-1}$, relative to the non-RLAGN galaxies should have changed from $z\approx1$ to $z\leq0.3$.

We notice that although the test for the second subsample ($0.6<z\leq1$, $10^{22.5}<L_{1.4\rm\,GHz}/({\rm W\,Hz^{-1}})\leq10^{23.5}$) gives a large $p$-value, which stands out among all four tests, it also gives an K-S statistic of 0.1, which is close to the K-S statistic in the test for the local RLAGN subsample.
The similar statistic but different $p$-value reflects that the K-S test is less sensitive in the high-redshift sample because of the smaller sample size.
Nevertheless, we consider that the K-S test should still be sensitive enough to probe the difference between the $q$ distributions of RLAGNs and control samples.
First, the three RLAGN subsamples from the COSMOS contain similar numbers of sources, but K-S tests can still give $p$-values close to or less than 1\% for the first and the third subsamples.
Second, to prove the K-S test is sensitive enough to probe a RLAGN-rounder trend that is seen in the fourth panel of Fig.\ref{fig:qdist} in a sample as small as the second subsample, we performed some simple tests.
For clarity, we note the second subsample as Sample H and the local subsample as Sample L.
We resampled the RLAGNs in Sample L to make a test RLAGN sample with a \mstar distribution and sample size similar to those of Sample H.
We also resampled the local non-RLAGN in the same manner to make a test non-RLAGN sample.
We then performed a similar control sample analysis using these two test samples to see if the K-S test can still work.
The results give a median $p$-value of 0.5\%.
Therefore, if the intrinsic difference between RLAGNs and non-RLAGNs in Sample H is as large as in Sample L, the K-S test result for Sample H should still give a $p$-value of less than 1\%, instead of 37.9\% as shown in the fourth panel of Fig. \ref{fig:qdist}.
In addition, we performed a resampling by selecting sources from Sample H to make test samples with \mstar distributions similar to that of Sample L.
We generated two kinds of test sample, one with a sample size similar to Sample H and the other with a sample size similar to Sample L.
We performed K-S tests to compare the $q$ distributions of these test RLAGNs and non-RLAGNs.
This resampling is to test whether a different \mstar distribution will bias the K-S results for comparing the $q$ distribution of RLAGNs and non-RLAGNs.
When the sample size is the same as Sample L, the K-S tests have a median $p$-value of less than $10^{-4}$.
This $p$-value goes up to about 10\% when the sample size is reduced to that of Sample H.
Therefore, changing the \mstar distributions does not lead to different significance levels in Sample H and Sample L.
The differences between $p$-values in the second and the fourth panels are not driven by the different \mstar distributions.
Therefore, the lack of significant difference between RLAGNs and non-RLAGNs in the second subsample cannot be explained purely by the small sample size or different \mstar distributions.

\subsection{RLAGN morphological changes at different redshifts}\label{sec:compare}
In the previous sections, we show that high-redshift RLAGNs have a different morphological distribution compared to the low-redshift RLAGNs.
To directly study the RLAGN morphological changes at different redshifts, we used the local RLAGN sample described in \citetalias{Zheng20} as a comparison.

This local sample is based on the LoTSS DR1 and has different distributions of \mstar\,and radio luminosity compared to the VLA-COSMOS data.
All the RLAGNs in the local sample have redshift lower than 0.3, while all the RLAGNs from the VLA-COSMOS data have redshifts higher than 0.3.
We show the RLAGNs in the two samples in the \mstar--$L_{150\rm\,MHz}$ plane in the left panel of Fig.\ref{fig:compare}, where the $L_{150\rm\,MHz}$ for sources in the VLA-COSMOS survey are derived assuming a canonical radio spectral index of 0.7.
The majority of the local RLAGN sample have a lower luminosity and slightly larger \mstar, but we can still find an overlapping region ($10^{11}<M_{\star}/M_{\odot}\leq10^{11.6}$, $10^{23}\rm\,W\,Hz^{-1}<L_{150\rm\,MHz}\leq10^{25.5}\rm\,W\,Hz^{-1}$) as shown in the left panel of Fig.\ref{fig:compare}, where both have a considerable fraction of sources.
We can therefore construct a local RLAGN control sample that is directly comparable with this work.

We divide the sampling region into 25 small cells as shown in Fig.\ref{fig:compare}, and then randomly select a sample of local RLAGNs with the same number of sources in each cell as the high-redshift sample.
In this way, we obtain a local RLAGN control sample with \mstar and radio luminosity distributions similar to those of the high-redshift sample in the sampling region.
The difference between the low- and high-redshift sample was then assessed using a K-S test.
These steps were also repeated 1000 times to reduce the stochastic error. 
The right panel of Fig.\ref{fig:compare} presents the resulting $q$ distribution of the control sample together with the median K-S statistic from 1000 tests.

From Fig.\ref{fig:compare}, it is apparent that the high-redshift RLAGN sample peaks at smaller $q$ than the local control sample.
The K-S test gives a statistic of 0.16 and a $p$-value $=2.83\%$, rejecting the null hypothesis at 2$\sigma$ level.
Therefore, the hosts of high-redshift RLAGNs typically have a smaller axis ratio than those of local RLAGNs with similar \mstar and radio luminosities, but the significance of this difference is not high.
We checked the possibility that the difference is due to bias in measurements and sample selections. 
However, we find that the systematic difference in $q$ measurements, the misclassification of RLAGNs, and the systematic difference in \mstar estimations would not  greatly change the significance of our results.

In conclusion, we suggest that the difference in $q$-distributions of the two samples is not likely caused by a bias in measurements or in sample selections.
The relatively small sample size may be a factor leading to the low significance.
In addition, the RLAGNs with redshift 0.3 to 0.6 show similar distribution in the $L-q$ plane with the local RLAGNs, as revealed in  Fig.\ref{fig:Lvsq}, which may also lower the significance in this analysis.
To better confirm this, a larger sample of high-redshift RLAGNs is needed in the future.

\section{Summary and Discussion}\label{sec:conclusion}

In this work, we used the radio data from the 3GHz VLA-COSMOS Large project and the morphology information listed in the ACS-GC for the COSMOS region to study the link between radio luminosity and host galaxy shape for RLAGNs with redshifts above 0.3 and $L_{1.4\rm\,GHz}$ larger than $10^{22}\rm\,W\,Hz^{-1}$. 
We constructed a mass-complete sample containing 3093 colour-selected passive galaxies, 234 of which host a RLAGN.
We investigated the distribution of RLAGNs in the $L_{1.4\rm\,GHz}-q$ plane and compared these RLAGNs with both non-RLAGNs and local RLAGNs using control samples.
Our main results can be summarised as follows:
\begin{itemize}
    \item Within redshifts 0.3 to 0.6, the radio luminosity of low-mass ($10^{10.8}-10^{11.2}\,M_{\odot}$) RLAGNs does not show a dependence on galaxy axis ratio, while the high-mass ($10^{11.2}-10^{11.5}\,M_{\odot}$) RLAGNs can reach a high luminosity only when they have a large axis ratio.
    \item For RLAGNs with redshifts of 0.6 to 1 and stellar mass from $10^{11.2}$ to $10^{11.5}\,M_{\odot}$, the radio luminosity does not depend on the host galaxy axis ratio.
    \item Both RLAGNs with $0.3<z\leq0.6$ and high-luminosity RLAGNs with $0.6<z\leq1$ have a significantly different $q$ distribution from those of the non-RLAGN \mstar-controlled comparison sample.
    RLAGNs are typically rounder than non-RLAGNs with similar \mstar.
    \item The $q$ distribution of low-luminosity ($L_{1.4\rm\,GHz}$ from $10^{22.5}$ to $10^{23.5}\rm\,W\,Hz^{-1}$) RLAGNs does not appear to be different from that of the non-RLAGN \mstar--controlled sample, in contrast to the low-redshift RLAGNs within this luminosity range.
    \item High-redshift RLAGNs are likely to have smaller $q$ than those of their local control samples with similar \mstar and radio luminosity distributions.
\end{itemize}

Based on the results above, the link between RLAGN radio luminosity and host galaxy shape at high redshift ($z>0.6$) seems to be different from that in the nearby Universe.
%
We find evidence suggesting that a higher fraction of high-luminosity RLAGNs reside in elongated galaxies at high redshift than at low redshift.
At $z<0.6$, as shown here and in previous work on RLAGNs with $z<0.3$ \citep[and  \citetalias{Zheng20}]{Barisic19}, low-mass and low-power RLAGNs span a wide range of $q$, while high-power RLAGNs typically all have large $q$.
Based on the spin paradigm \citep{Blandford77,Wilson95,Fanidakis11}, in which black hole spin determines the ability to launch a radio jet, the difference in the host galaxy shapes of high- and low-power RLAGNs at low-redshift can be explained by two spin-up paths: low-power sources are spun up via secular processes while high-power sources are spun up via major mergers, and therefore disc-like galaxies are not likely to have AGNs with high radio luminosities.
{ Assuming this picture is also true at high redshift, the presence of high-power RLAGNs with small $q$ implies that either a fraction of high-power RLAGN hosts today had an elongated shape in the past, or that local sources spun up via secular processes were more luminous at high redshift.

It is not surprising that these hosts of high-power small-$q$ RLAGNs have evolved into round elliptical galaxies today.
Observations and simulations have shown that galaxies with \mstar$>10^{10}\,M_{\odot}$ have undergone an average of between one and two major mergers since $z=1.2$ and the fraction of spheroid-like galaxies has increased significantly \citep{Chang13,Lotz08,Conselice09b,Conselice14}.
However, this does not explain why these small-$q$ RLAGNs have such high radio luminosities at high redshift.

If these elongated galaxies had followed a similar formation path to that of the elongated galaxies today, the spin of SMBHs in low- and high-redshift galaxies would also be similar. 
Therefore, the existence of high-power small-$q$ RLAGNs implies that small-$q$ RLAGNs have a higher accretion rate or a different conversion factor between jet power and radio luminosity at high redshift.
This is consistent with the trend whereby, within $z<1$, the earlier Universe has a higher density and gas fraction \citep[e.g.]{Decarli20}.
With a higher gas fraction, it is easier for SMBHs to reach a high accretion rate and jet power.
A denser environment could also lead to a more confined radio lobe and less energy loss from adiabatic expansion, such that a higher radio luminosity can be achieved at a given jet power \citep{Barthel96}.
The consequence of these effects is that some elongated galaxies, which can only have a weak jet in the local Universe, can be more luminous at high redshift, and therefore the RLAGNs at a given radio luminosity and \mstar will contain more elongated sources.

The luminosity evolution means that, if a high-redshift RLAGN was moved to the local Universe, it would become less powerful, and therefore the counterpart of a high-redshift RLAGN could be a local RLAGN with lower luminosity.
Therefore, we can estimate the influence of this luminosity evolution by shifting the sampling region of local RLAGNs in the control sample analysis.
Assuming high-redshift RLAGNs are on average brighter than low-redshift RLAGNs by $\Delta{\rm log}\, L_{\rm 150\rm\,MHz}$, we shifted the sampling regions of the LoTSS RLAGN in Sect. \ref{sec:compare} and constructed control samples for VLA-COSMOS RLAGN with \mstar$=10^{11.12}-10^{11.48}\,M_{\odot}$, $L_{150\rm\,MHz}=10^{23.5}-10^{25}\rm\,W\,Hz^{-1}$.
We obtain the null hypothesis probability as a function of $\Delta{\rm log}\, L_{\rm 150\rm\,MHz}$ from K-S tests comparing axis ratio distributions of VLA-COSMOS RLAGNs with the control sample at different $\Delta{\rm log}\, L_{\rm 150\rm\,MHz}$.
At every 0.1 dex step in luminosity difference, from 0 to 1 dex, we performed 1000 simulations; we plot the results in Fig. \ref{fig:loffset}.
The $p$-value begins to exceed the 5\% threshold at $\sim$0.3 dex, and at $\sim$0.6 dex less than half of the simulations reject the null hypothesis and the $p$-value begins to increase quickly.
This 0.3-0.6 dex luminosity offset is in line with the average radio luminosity difference between $z=0.14$ and $z=0.55$ at \mstar$=10^{11.2}-10^{11.5}\,M_{\odot}$ \citep{Donoso09}.
Therefore, the evolution of the shape of RLAGN can be interpreted as a byproduct of luminosity evolution.
This means that the influence of galaxy morphology on RLAGNs should not change significantly over redshifts 0 to 1 based on current data.

However, it should be noted that the direct comparison between high- and low-redshift RLAGNs is only constrained in a small stellar mass range.
According to the analysis of low-redshift RLAGNs, the different correlations between radio luminosity and galaxy morphology become most significant for very massive galaxies \citepalias{Zheng20}.
Therefore, to  study this further, large samples containing a considerable fraction of more radio-luminous AGNs ($L_{\rm150\,MHz}\gtrsim10^{24.6}\,\rm W\,Hz^{-1}$) are needed.
\begin{figure}
    \centering
    \includegraphics[width=\linewidth]{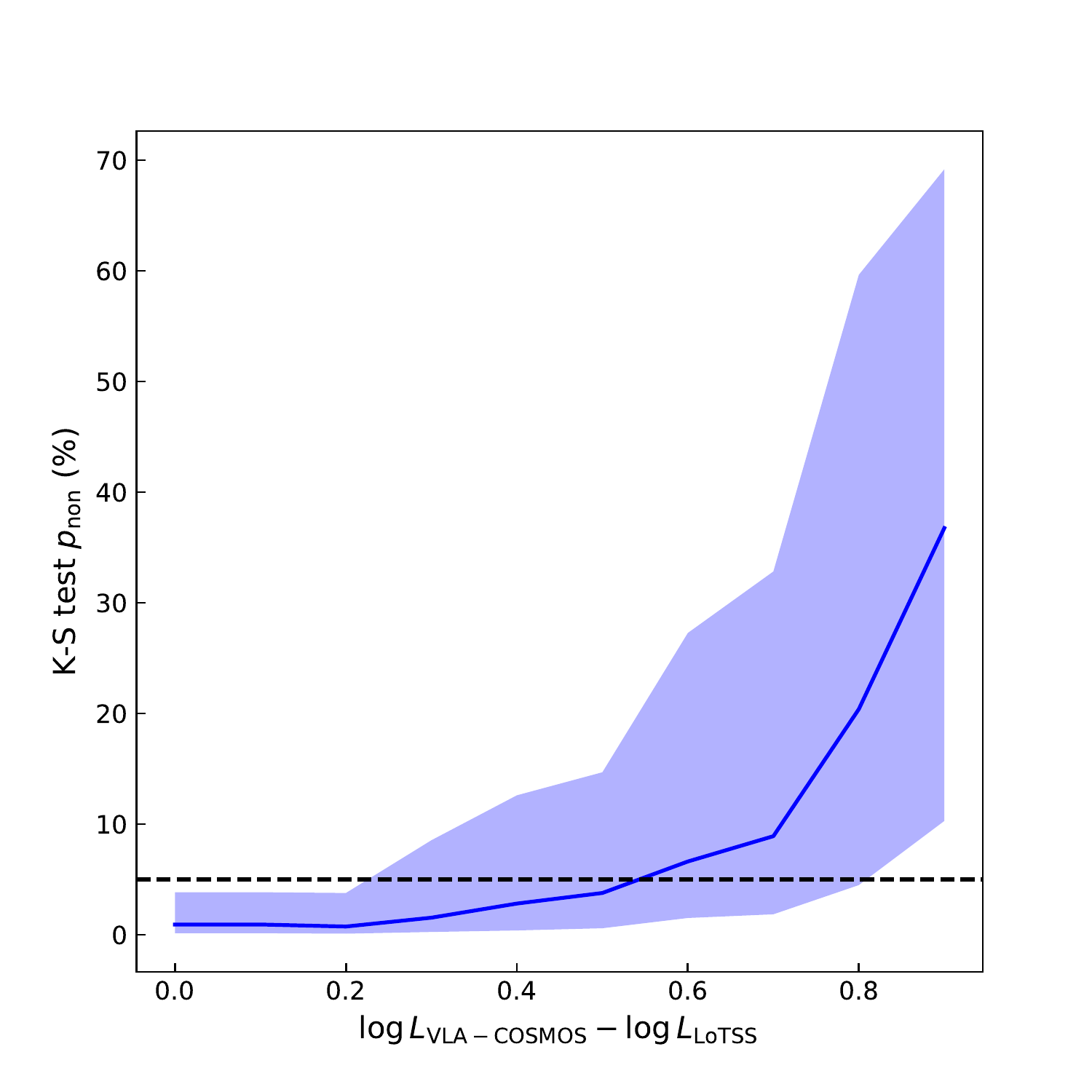}
    \caption{
    Null hypothesis probability as a function of luminosity difference between high-redshift and low-redshift RLAGNs at fixed stellar mass.
    The blue line denotes the median $p$-value from 1000 simulations at different $\Delta{\rm log}\, L_{\rm 150\rm\,MHz}$, while the shaded area represents the 1$\sigma$ deviations.
    The black dashed line represents the 5\% threshold below which the null hypothesis is rejected.
    }
    \label{fig:loffset}
\end{figure}

We note that all high-power small-$q$ RLAGNs ($q\leq0.7$ and $L_{1.4\,\rm GHz}>10^{23.5}\,\rm W\,Hz^{-1}$ in Fig.\ref{fig:Lvsq}) at high redshift are identified as low-to-intermediate radiative luminosity AGNs in \citet{Smolcic17b}.
Based on the analysis of \citet{Smolcic17b} and \citet{Delvecchio17}, this means that they are all more likely to be jet-mode AGNs, which typically have a low Eddington-scaled accretion rate \citep[$<10^{-2}$; ][]{Best12}.
Therefore, their accretion modes are similar to the RLAGNs in massive elliptical galaxies, which usually have a large $q$.
This is consistent with the local low-luminosity RLAGN results in \citetalias{Zheng20}, where the authors  found that morphology type is not likely to be analogous to the accretion-based dichotomy.

It is also interesting to investigate whether compact steep-spectrum (CSS) and peaked-spectrum (PS) sources have an impact on our results. 
These sources have a convex radio spectrum that peaks at $\sim$0.1 to 5 GHz and  are therefore easier to detect in the 3GHz survey. 
There is evidence that these sources have higher SFRs and denser interstellar gas \citep[see][and reference therein]{ODea21} as compared to the normal spectrum population.
This means that they are also more likely to be discy.
However the fraction of CSS and PS sources in our samples should be small. 
Previous surveys with a flux density limit of 0.1$\sim$1 Jy have a CSS and PS source fraction of the order of 10\% and the fractions seem to decrease at lower flux densities \citep{Snellen00,Callingham17,ODea21}.
Therefore, at the faint flux limits of the LoTSS sample and the VLA-COSMOS \citep{Shimwell19,Smolcic17a}, the potential bias of CSS and PS sources on the results presented is likely to be small. 

Our work suggests that the host galaxies of high-redshift RLAGNs have a different morphology distribution from those of low-redshift RLAGNs.
The most straightforward interpretation of this evolution is that the link between the radio luminosity of a RLAGN and the shape of its host galaxy is in fact the same at high and low redshift, but that higher redshift RLAGNs are more luminous by a factor of 2 to 4. 
To further study the intrinsic morphological evolution of RLAGNs and address the uncertainties in this work, it would be necessary to investigate a large sample of massive galaxies with radio-luminous AGNs.  
This is likely to be achieved with the help of larger deep-field radio surveys such as the LoTSS Deep Fields.}

\begin{acknowledgements}
     We thank Emma Rigby for checking and polishing the writing. X. C. Z. acknowledges support from the CSC (China Scholarship Council)-Leiden University joint scholarship program. KJD acknowledges support from the European Union’s Horizon 2020 research and innovation programme under the Marie Skłodowska-Curie grant agreement No. 892117.
\end{acknowledgements}

%
%

\bibliographystyle{aa}
\bibliography{HzRLshape}

\begin{appendix}

\section{Consistency of axis ratio measurements}\label{app:q}
\begin{figure*}
\centering
    \includegraphics[width=\linewidth]{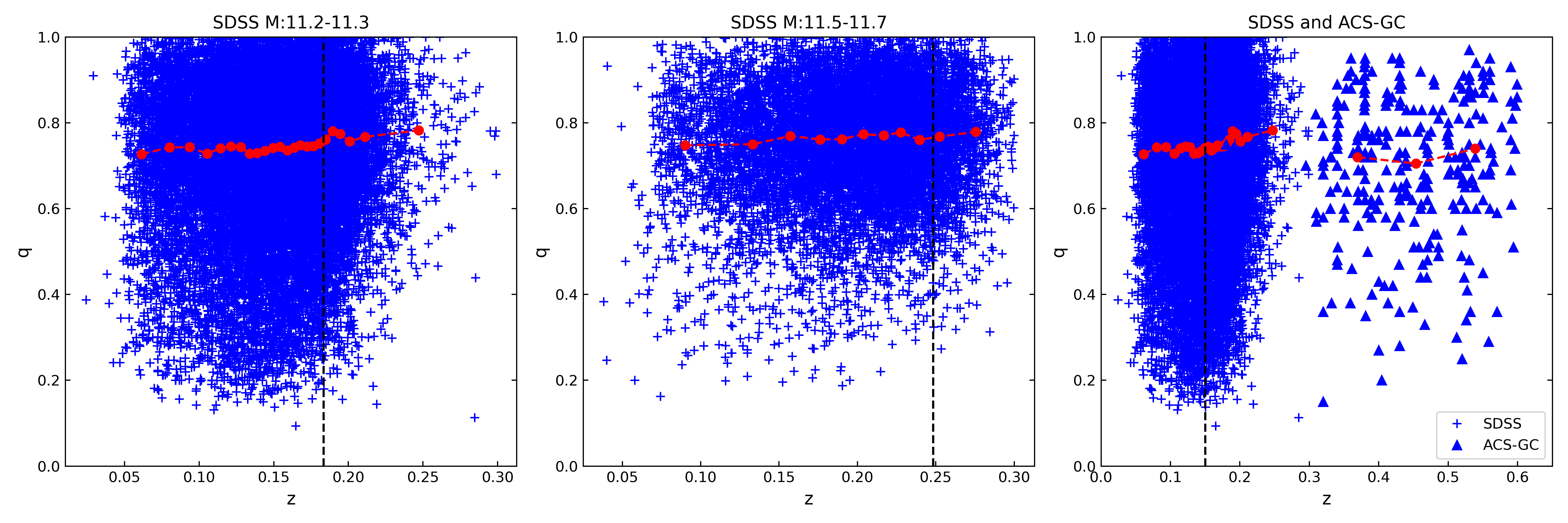}
    \caption{
    Axis ratios of galaxies as a function of redshifts in different samples.
    Left: Axis ratio of passive galaxies with \mstar=$10^{11.2}-10^{11.3}M_{\odot}$ from the SDSS as a function of redshift.
    The red dots are the median for every 1000 sources.
    The black dash line denotes the mass-complete redshift.
    Middle: Axis ratio of passive galaxies with \mstar$=10^{11.5}-10^{11.7}M_{\odot}$ from the SDSS.
    Right: Axis ratio of the passive galaxies from the SDSS and the ACS-GC.
    The galaxies from SDSS are with \mstar=$10^{11.2}-10^{11.3}M_{\odot}$ and marked with blue crosses, while the galaxies from the ACS-GC are with \mstar=$10^{11}-10^{11.5}M_{\odot}$ and marked with blue triangles.
    The median $q$ of sources from the ACS-GC are calculated for every 100 sources.
    }
    \label{fig:sdssab}
\end{figure*}
To investigate the consistency of axis ratio measurements in the SDSS and the ACS-GC, and particularly to check whether or not the lower spatial resolution of SDSS could introduce bias into the analysis, we selected two galaxy samples from the SDSS DR16 and inspected the variation of their axis ratio as a function of redshift.
Both samples are selected to be passive galaxies with ranges of redshift and $r$ band magnitude similar to that of \citetalias{Zheng20}: $0.01<z\leq0.3$, $14.5<r\leq17.77$, and $u-r>1.6\times(r-z)+1.1$.
One of the samples has stellar mass within $10^{11.2}-10^{11.3}\,M_{\odot}$, and the other one has stellar mass within $10^{11.5}-10^{11.7}\,M_{\odot}$.

The axis ratios of galaxies in these sample as a function of redshifts are shown in Fig.\ref{fig:sdssab}.
We also show the median axis ratio for every 1000 sources and the mass-complete redshift based on the mass completeness limit used in \citet{Chang15} and \citetalias{Zheng20}.
The galaxies at these \mstar ranges are complete only below the mass-complete redshift.

Apparently, the median $q$ for galaxies below the mass-complete redshift does not vary significantly despite the fact that the spatial resolution drops significantly from $\sim1.2$kpc at $z\sim0.05$ to $\sim4.7$kpc at $z=0.25$, assuming a typical angular resolution of 1.2$\arcsec$.
This means that the $q$ measurement from the SDSS are consistent and reliable at these redshifts.
The median $q$ shows a slight increase only at redshifts above the mass-complete redshift.
This is mainly because of the lack of some low-luminosity flat galaxies at these redshifts, which can also be seen from Fig.\ref{fig:sdssab}.
In our works, these galaxies are discarded in the sample-selection process.

To compare the measurements from the SDSS and the ACS-HST, we also chose a mass-complete passive galaxy sample at $0.3<z\leq0.6$  from the ACS-GC to compare with the first SDSS sample.
The galaxies are selected as in Sect.\ref{sec:data} in this work.
The spatial resolution of this sample is nearly the best in our work ($\sim0.45$kpc at $z=0.3$).
The \mstar range of this sample is $10^{11}-10^{11.5}M_{\odot}$, which is not identical to the first SDSS sample, because there are few sources in this sample within $10^{11.2}M_{\odot}$ and $10^{11.3}M_{\odot}$ and the median $q$ for sources from the SDSS does not change significantly in this \mstar range.
We calculated the median axis ratio for every 100 sources from the ACS-GC sample and show them in the right panel of Fig.\ref{fig:sdssab}.

From the right panel of Fig.\ref{fig:sdssab}, we can see that the median axis ratio from the ACS-GC sample is similar to that from the SDSS sample.
Therefore, we conclude that the $q$ measurements from the SDSS and the ACS-GC are consistent.
\end{appendix}

\end{document}